\documentclass[prb,two column,floats,showpacs,amsmath,amssymb,superscriptaddress]{revtex4}

\usepackage{graphicx}
\usepackage{array}
\usepackage{hyperref}

\usepackage{color}

\newcommand{\I}{\mathrm{i}}
\newcommand{\av}[1]{\langle #1 \rangle}
\newcommand{\bra}[1]{\left\langle #1 \right|}
\newcommand{\ket}[1]{\left| #1 \right\rangle}

\newcommand{\ua}{\uparrow}
\newcommand{\da}{\downarrow}

\newcommand{\szlta}{\av{S_{z}}_{T}}

\begin{document}

% -------------------------------------------------------------------------------------------------------------------------------------------
% document settings
% -------------------------------------------------------------------------------------------------------------------------------------------

% -------------------------------------------------------------------------------------------------------------------------------------------
% beginning of text-body
% -------------------------------------------------------------------------------------------------------------------------------------------

\title{Ultra long spin decoherence times in graphene quantum dots with a small number of nuclear spins}

\author{Moritz Fuchs}
\affiliation{Institut f\"{u}r Theoretische Physik und Astrophysik,
Universit\"{a}t W\"urzburg, D-97074 W\"urzburg, Germany}

\author{John Schliemann}
\affiliation{Institut f\"{u}r Theoretische Physik,
Universit\"{a}t Regensburg, D-93053 Regensburg, Germany}

\author{Bj\"orn Trauzettel}
\affiliation{Institut f\"{u}r Theoretische Physik und Astrophysik,
Universit\"{a}t W\"urzburg, D-97074 W\"urzburg, Germany}

\date{\today}

\begin{abstract}
We study the dynamics of an electron spin in a graphene quantum dot, which is interacting with a bath of less than ten nuclear spins via the anisotropic hyperfine interaction. Due to substantial progress in the fabrication of graphene quantum dots, the consideration of such a small number of nuclear spins is experimentally relevant. This choice allows us to use exact diagonalization to calculate the long-time average of the electron spin as well as its decoherence time. We investigate the dependence of spin observables on the initial states of nuclear spins and on the position of nuclear spins in the quantum dot. Moreover, we analyze the effects of the anisotropy of the hyperfine interaction for different orientations of the spin quantization axis with respect to the graphene plane. Interestingly, we then predict remarkable long decoherence times of more than 10ms in the limit of few nuclear spins.
\end{abstract}

\pacs{76.20.+q, 76.60.Es, 85.35.Be, 03.65.Yz, 81.05.ue}

\maketitle

\section{Introduction}
\label{sec:introduction}

In recent years, spin qubits hosted in solid state nanostructures have been under extensive research due to their possible applications in quantum information processing and computation. Among the host materials of spin qubits, quite different approaches can be found, for instance, III-V-semiconductor and carbon nanotube quantum dots \cite{Hanson2007} (QD) as well as nitrogen vacancies in diamond \cite{Doherty2013}. These host materials show promising prospects but, unfortunately, also come with certain drawbacks.

A precise control of the qubit state is the major advantage of III-V-semiconductor QDs based on Al(Ga)As heterostructures. Preparation and readout of the qubit with high fidelity via electrostatical gates has been demonstrated in many ground-breaking experiments\cite{Petta2005,Koppens2008,Greilich2009,Eble2009,Foletti2009,Ladd2010,Press2010,Bluhm2010,Bluhm2010a,Shulman2012,Dial2013}. However, the disadvantage of this material system is the presence of many nuclear spins inherent to the atoms of group III and group V elements of the periodic table. These nuclear spins give rise to a fast decoherence of the electron spin\cite{Schliemann2003,Coish2009}. Elaborate design of experiments including pulse sequences and methods to polarize the nuclear spins\cite{Reilly2008,Xu2009,Latta2009,Vink2009,Issler2010,Carter2011} such as dynamic nuclear polarization may help to compensate the effect of the spin bath. Nevertheless, it seems desirable to reduce the number of nuclear spins  which can be achieved on the basis of other host materials.

Obvious candidates are carbon and silicon, since their spin carrying isotopes have only very low natural abundances of about $1\%$ and $5\%$, respectively. In silicon, the qubits can be fabricated\cite{Morello2010,Morton2011} either using donor impurities or by confining a single electron via electrostatical gates. However, a controlled localization of the donor impurities is still a challenging task and electrostatically confined Si QDs often involve nanostructures with other materials like Ge, which potentially introduce additional nuclear spins\cite{Morton2011}. Carbon based QDs can be realized by confining an electron spin in carbon nanotubes\cite{Kuemmeth2008,Churchill2009,Churchill2009a,Steele2009,Monge2011,Chorley2011,Eichler2011,Lai2012} (CNT) via electrical gates allowing for a good control in the few electron regime. However, the curvature of the CNTs gives rise to a sizable spin-orbit coupling, yet another intrinsic source of decoherence to the electron spin. A different approach to a carbon based QD is the use of nitrogen vacancies in diamonds\cite{Dutt2007,Neumann2008,Togan2010,Togan2011,Doherty2013}, which show tremendously long coherence times. Unfortunately, control and readout of the qubit have to be done optically, which is disadvantageous for the realization of future on-chip electric circuits.

These examples illustrate a more general issue of designing qubits, where an easy (electric) control and scalability seem to compete with noiseless environments and, hence, long decoherence times. A system potentially providing the best of both worlds is a graphene QD \cite{Trauzettel2007,Recher2010}, which offers very interesting electronic properties\cite{CastroNeto2009} and a small spin-orbit coupling\cite{Huertas-Hernando2006,Min2006,Yao2007,Gmitra2009}, as well as the possibility to control the number of nuclear spins by isotopic purification\cite{Banholzer1992,Simon2005,Churchill2009}. Moreover, the hyperfine interaction between the remaining nuclear spins and the electron spin is much smaller than in GaAs or Si. Additionally, the hyperfine interaction in graphene is anisotropic which could provide interesting applications as we discuss at the end of this article.

Experimentally, graphene QDs are, for instance, realized by confining electrons with gates in bilayer graphene\cite{Goossens2012,Allen2012} and graphene nanoribbons\cite{Liu2009,Liu2010a}, respectively, or by etching the QD structure out of graphene flakes\cite{Stampfer2008,Ponomarenko2008,Schnez2009,Molitor2009,Guttinger2009,Molitor2010,Wang2010,Guttinger2010,Guttinger2011,Fringes2011,Engels2013}. Typical diameters are of the order of tens to hundreds of nanometers resulting in $K=15$ to $K=1500$ nuclear spins assuming a natural abundance of spin carrying $^{13}{\rm C}$ of $1\%$. Thus reducing the abundance of $^{13}{\rm C}$ by only two orders of magnitude leads to very small spin baths even in the case of rather large QDs. Recently, ultra small graphene QDs with diameters in the $1\,\mathrm{nm}$ range were made by electroburning\cite{Barreiro2012}. Altogether, these considerations show that the study of few nuclear spin models with $K<10$ as considered in this work is highly relevant for future research in the field.

In this paper, we aim to set the basis for forthcoming investigations of the spin dynamics in graphene nanostructures. Besides quantum information theory, especially ongoing research on magnetism on edges\cite{Luitz2011,Schmidt2012,Schmidt2013,Golor2013,Golor2013a,Golor2013b} and vacancies\cite{Chen2011,Nair2012} in graphene can benefit from a detailed knowledge of the properties of the anisotropic hyperfine interaction (AHI). Moreover, we intend to complement our previous analytic study\cite{Fuchs2012} of the electron spin dynamics. Considering a large nuclear spin bath, we investigated the coherence of the electron spin in a non-Markovian approach using a generalized master equation. In this work, however, we were limited to large external magnetic fields in order to justify the perturbative treatment of the hyperfine interaction.

Since we restrict ourselves to less than ten nuclear spins in the present work, we can apply exact diagonalization to the hyperfine Hamiltonian, which offers a powerful tool to investigate the dynamics of the electron spin for a wide parameter regime\cite{Schliemann2002,Schliemann2003,Dobrovitski2003,Shenvi2005,Zhang2007,Zhang2008,Sarkka2008,Cywinski2010,Erbe2012}. In particular, we analyze the role of the number of nuclear spins $K$, their position within the QD, as well as their initial spin state. Thereby, we use the long-time average $\av{S_{z}}_{T}$ of the longitudinal electron spin and its decoherence time $T_{D}$ to quantify the influence of these different aspects. Moreover, we investigate the dependence of $\av{S_{z}}_{T}$ and $T_{D}$ on the orientation of the spin quantization axis with respect to the graphene plane. For the long-time average, we find a continuous crossover from a initial state dominated regime for $K<5$ to a regime more affected by the configuration of the nuclear spins for $K>6$ where the relative positions of the nuclear spins with respect to each other matter. As we will show below, this behavior can be understood by an analysis of the Hilbert space dimensions as well as of the spatial distribution of the nuclear spins in the QD. Besides this regime change, a growing nuclear spin bath suppresses fluctuations around the long-time average more and more effectively, while the average itself is almost constant for all $K<9$ with $\av{S_{z}}_{T}\approx \hbar/4$ in the out-of-plane orientation and $\av{S_{z}}_{T}\approx0$ in the in-plane case. By resolving the orientation dependence in more detail, we find good agreement with $\av{S_{z}}_{T}(\beta)=\av{S_{z}}_{T}(0)\cdot\cos(\beta)^{2}$, where $\beta=0$ and $\beta=\pi/2$  correspond to the out-of-plane and in-plane orientation, respectively, {\it cf.} Fig. \ref{fig:QD} (a).

Evidently, the decoherence time $T_{D}$ strongly depends on the number of nuclear spins $K$. We observe that the configuration of the nuclear spins is decisive even for very small numbers of nuclear spins, whereas the initial states play only a minor role. Depending on the relative positions of the nuclei, the decoherence times may deviate over several orders of magnitude. This behavior can be traced back to changes of the spectrum of eigenvalues of the full Hamiltonian. Moreover, the decoherence times significantly differ between out-of-plane and in-plane orientation. For $K=3$ and $\beta=0$, the majority of investigated configurations show very long decoherence times above $10\,\mathrm{ms}$ where, in many cases, even no decoherence at all was found. For $\beta=\pi/2$, in contrast, we always find decoherence, which predominantly occurs within $500\,\mu\mathrm{s}$. With increasing bath size, the decoherence times decrease for both orientations of the quantization axis. Then, decoherence times below $500\,\mu\mathrm{s}$ are most common.

The article is organized as follows. In Sec.~\ref{sec:model}, we explain our model of the QD and discuss all relevant interactions of the spins with each other. Subsequently, in Sec.~\ref{sec:method}, we present the method used to obtain both the long-time average of the electron spin and its decoherence times. All results are shown and analyzed in Sec.~\ref{sec:results}. Based on a summary in Sec.~\ref{sec:conclusion}, we give an outlook on possible applications of few nuclear spin graphene QDs and on interesting future projects in this field.

\section{Model}
\label{sec:model}

\begin{center}
  \begin{figure*}
  \centering
  \includegraphics[width=\textwidth,type=eps,ext=.eps,read=.eps]{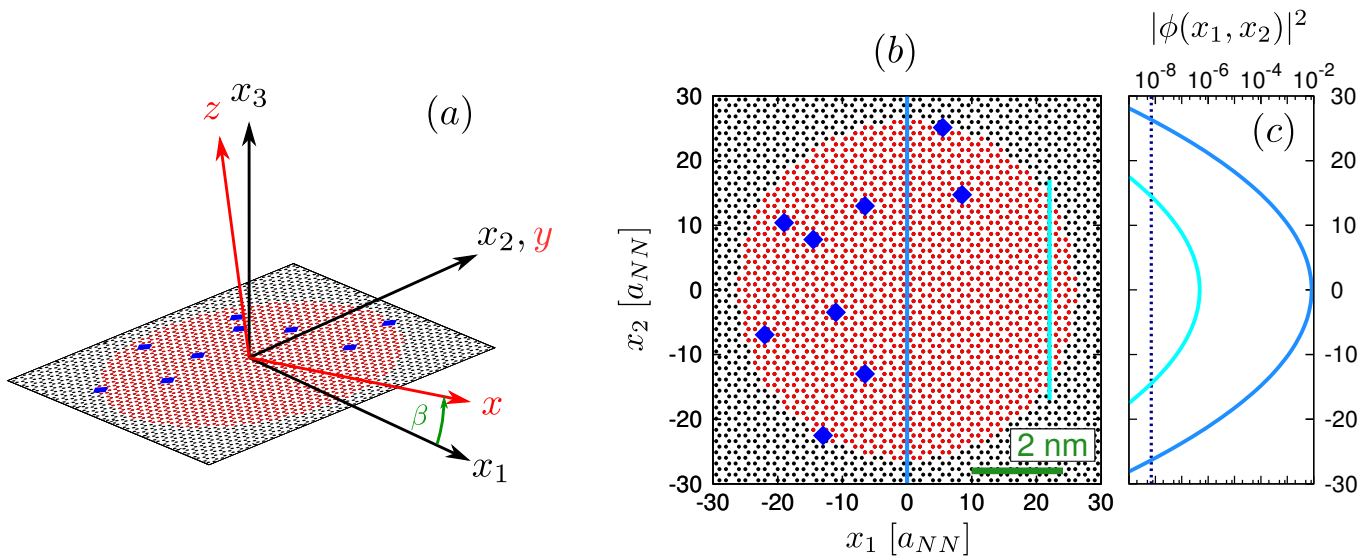}
  \caption{(Color online) \textbf{(a):} The graphene ($(x_{1},x_{2},x_{3})$) and quantization axis ($(x,y,z)$) reference frames. \textbf{(b):} A graphene QD (red sites) for a Gaussian envelope function with $K=10$ uniformly random distributed $^{13}{\rm C}$ atoms (blue squares) carrying a nuclear spin $1/2$. The extent of the dot over the graphene lattice is defined via the electron envelope function, where all sites within the dot obey the cut-off relation defined in Eq. (\ref{eq:QD_def}). \textbf{(c):} The envelope function for fixed $x_{1}=0$ and $x_{1}=22\,a_{NN}$, respectively. The dashed line indicates the cut-off defined in Eq. (\ref{eq:QD_def}).}
  \label{fig:QD}
  \end{figure*}
\end{center}

We study the spin dynamics in a graphene quantum dot, where one electron spin is in contact with a bath of nuclear spins hosted by the $^{13}{\rm C}$ atoms. Due to the confinement, the electron can occupy a discrete spectrum of bound states, with an energy splitting between different states\cite{Silvestrov2007,DeMartino2007,Trauzettel2007,Matulis2008,Recher2009,Titov2010}. If the temperature is small compared to the level spacing $\Delta E$ of these bound-state energies, the electron resides in the ground state, which we describe by an envelope function $\phi(\vec{r})$. Hence, the probability to find the electron in a certain region of the QD can be described by its absolute square $|\phi(\vec{r})|^2$. In this paper we define the \textit{``center``} of the dot as the region around $\vec{r}_{max}$, where the envelope function is maximal $|\phi(\vec{r}_{max})|^2$. Far away from this center, the envelope function has to vanish:
\begin{equation}
|\phi(\vec{r}_{max}+\Delta\vec{r})|^2\rightarrow0 \;\; \mathrm{for} \;\; |\Delta\vec{r}| \rightarrow \infty
\,.
\label{eq:integrability}
\end{equation}
In this work, we model a graphene QD by the set of atomic sites $\lbrace\vec{r}_{k}\rbrace_{i=1}^{N_{sites}}$ obeying
\begin{equation}
\frac{|\phi(\vec{r}_{i})|^2}{|\phi(\vec{r}_{max})|^2}>C
\,,
\label{eq:QD_def}
\end{equation}
where $C=10^{-6}$ is a constant cut-off. A plot of a QD realized in this way is shown in Fig. \ref{fig:QD}. The choice of this finite system of discrete sites $\vec{r}_{i}$ imposes a normalization condition
\begin{equation}
\sum\limits_{i=1}^{N_{sites}} |\phi(\vec{r}_{i})|^2 = 1
\,,
\label{eq:QD_norm}
\end{equation}
since we want to find the electron with probability $1$ somewhere within the dot. Effectively, we ignore everything outside the barrier defined by the cut-off, which is justified by the vanishing probability to find the electron there. This choice will become more clear when we discuss the hyperfine interaction between the electron and a single nuclear spin below.

A possible choice for the envelope function is a Gaussian
\begin{equation}
\phi\left(\vec{r}\right)
=
\phi_{0}\exp\left[-\frac{1}{2}\left( \frac{r}{R}\right)^{2}\right]
=
\phi\left(r\right)
\,,
\label{eq:envfunc}
\end{equation}
where $r=\left|\vec{r}\right|$ is the absolute value of the electron position and the norm $\phi_{0}$ is chosen to satisfy the normalization condition in Eq. (\ref{eq:QD_norm}). This assumption is also in agreement with a recent experiment investigating the wave function of a graphene QD with soft confinement\cite{Subramaniam2012}. Note that the envelope function is not the exact electron wave function, but should give a good approximation to the precise solution. This can be seen, for instance, in graphene QDs based on semi-conducting armchair nano-ribbons\cite{Trauzettel2007}. The most important aspects, which are captured by this specific choice, are the absence of nodes in the ground state, a peak of the wave function in the center as well as a strong decay at the edges of the dot, which is illustrated by Fig. \ref{fig:QD} (c).

Having defined the shape of our dot, we are now able to introduce the nuclear spins present in the system. Since we do not have further knowledge about the distribution of $^{13}{\rm C}$ within the dot, we randomly place the nuclear spins on the sites defined by Eq. (\ref{eq:QD_def}), where each site is chosen with equal probability. An example of a configuration of ten nuclear spins is shown in Fig. \ref{fig:QD}.

We now proceed to the interactions between the nuclear spins and the electron spin and between themselves. The most relevant spin-spin interaction in our system is the hyperfine interaction between the electron spin $\vec{S}$ and $K$ nuclear spins $\vec{I}_{k}$ located at sites $\vec{r}_{k}$,
\begin{equation}
\hat{H}_{HI}
=
A_{HI} \sum\limits_{k=1}^{K} \sum\limits_{\mu,\nu} \overleftrightarrow{A}_{\mu\nu}\, |\phi(\vec{r}_{k})|^{2}\,\hat{S}_{\mu}\hat{I}_{k,\nu}
\,,
\label{eq:ahi}
\end{equation}
where the indices $\mu$ and $\nu$ run over spatial coordinates $x,y,z$. The energy scale of this interaction is given by $A_{HI}=0.6\,\mu\mathrm{eV}$ and $\overleftrightarrow{A}_{\mu\nu}$ is a spherical tensor\cite{Sakurai2010} of rank $2$, which takes into account the anisotropy of the hyperfine interaction in graphene\cite{Fischer2009a}. Remarkably, this interaction is strongly modulated by the envelope function, a fact which arises from the on-site nature of the hyperfine interaction. Thus, making the boundary of the dot smooth by taking a very small cutoff $C\rightarrow0$ in Eq. (\ref{eq:QD_def}), would only add vanishingly small contributions to the interaction in Eq. (\ref{eq:ahi}). Therefore we choose a small, but finite cutoff for simplicity.

Besides the anisotropic hyperfine interaction (AHI), there is also a dipole-dipole interaction between pairs of nuclear spins. In the parameter regime considered in this work, this interaction, however, is about five orders of magnitude smaller than the AHI and, thus, neglected. We also proved its irrelevance by a numerical study, which we will not present here.

Since external magnetic fields allow to manipulate spins experimentally, we include a Zeeman-Hamiltonian to account for this:
\begin{equation}
\hat{H}_{ZE}
=
\hbar \gamma_{S} B_{z} \hat{S}_{z} + \hbar \gamma_{^{13C}} B_{z} \sum\limits_{k=1}^{K} \hat{I}_{k,z}
\approx
A_{ZE} \hat{S}_{z}
\,,
\label{eq:aeb}
\end{equation}
where we used the fact that the electron gyromagnetic ratio $\gamma_{S}=1.76\times10^{11}\mathrm{s}^{-1}\mathrm{T}^{-1}$ is much larger than the gyromagnetic ratio of the nuclear spins $\gamma_{^{13}{\rm C}}=6.73\times10^{7}\mathrm{s}^{-1}\mathrm{T}^{-1}$ to justify the right-hand side of Eq. (\ref{eq:aeb}).

In the presence of an external magnetic field, an interplay of the spin orbit coupling with acoustic phonons can lead to spin relaxation times $T_{1}$ ranging from milliseconds to seconds\cite{Struck2010,Droth2011,Droth2013,Hachiya2013} for small external magnetic fields. The exact value of $T_{1}$ significantly varies with the spectrum of the phonons which depends on the details of the dot nanostructure. Providing that the graphene flake is flat throughout the spatial extent of the QD, however, the spin orbit interaction should be small. Thus, this assumption justifies to neglect the influence of the spin orbit interaction on our problem.

In the following, we aim to simulate a model experiment consisting of a preparation of the spins and the actual measurement of the spin dynamics. For the preparation, one can think of two different scenarios. First, the states of both the electron spin and the nuclear spins can be prepared in the presence of a strong external magnetic field $B_{0}\gg K \cdot A_{HI}|\phi(r_{max})|^2/\hbar\gamma_{S}=K\cdot|\phi(r_{max})|^2\cdot5.7\,\mathrm{mT}$, which imprints a well defined quantization axis. At the beginning of the actual measurement, this field is turned off or reduced to a finite value and the time evolution of the quantity of interest is recorded. However, since the tuning of magnetic fields is typically slow, this preparation scheme might not be adequate for a real experiment and we have to look for other solutions. In this case, we can think of injecting a spin polarized electron via a spin dependent tunneling process from a normal lead or via a spin conserving tunneling process from a spin polarized lead.

Anyhow, in both considered scenarios our system features two natural reference frames, the first defined by the graphene plane and the second one by the quantization axis of the electron as depicted in Fig. \ref{fig:QD} (a). In order to clarify the notation, the graphene coordinate system (GCS) is written as $\vec{\tilde{v}}=(\tilde{v}_{x_{1}},\tilde{v}_{x_{2}},\tilde{v}_{x_{3}})$ with all objects being marked with a tilde, whereas a vector in the quantization axis coordinate system (QCS) is labeled by $\vec{v}=(v_{x},v_{y},v_{z})$ \textit{cf.} Fig. \ref{fig:QD}. If one chooses, without loss of generality, the $x_{2}$- and the $y$-axis to coincide, both coordinate systems are connected via a rotation $\hat{D}(\beta)$ by an angle $\beta$ around this common axis.

Due to the symmetries of the carbon orbitals \cite{Fischer2009a}, the spherical tensor $\overleftrightarrow{A}_{\mu\nu}$ of the AHI in Eq. (\ref{eq:ahi}) takes its simplest form in the GCS, namely
\begin{equation}
\overleftrightarrow{\tilde{A}}
=
\left(
  \begin{array}{ccc}
  -\frac{1}{2}	&	0		&	0	\\
  0		&	-\frac{1}{2}	&	0	\\
  0		&	0		&	1
  \end{array}
\right)
\,,
\label{eq:spherical_tensor}
\end{equation}
while the spin operators, $\hat{S}_{\mu}$ and $\hat{I}_{k,\nu}$, and the spin states are most conveniently defined in the QCS.

In the main part of this work, we are interested in the time-dependent expectation value of an arbitrary operator $\hat{O}$
\begin{equation}
\av{O}(t)
=
\bra{\psi_{0}}\,\hat{U}^{\dagger}(t)\,\hat{O}\,\hat{U}(t)\,\ket{\psi_{0}}
\,,
\label{eq:av_operator}
\end{equation}
where $\ket{\psi_{0}}$ describes the initial state of the total spin system and $\hat{U}(t)=\exp[-\I\hbar^{-1}t\,\hat{H}]$ is the time evolution operator determined by the total Hamiltonian $\hat{H}=\hat{H}_{HI}+\hat{H}_{ZE}$. Note, that this is the total Hamiltonian with respect to the QCS, where the Zeeman Hamiltonian is always diagonal and the AHI Hamiltonian is obtained from its simple GCS form $\hat{\tilde{H}}_{HI}$ by
\begin{equation}
\hat{H}_{HI}
=
\hat{D}(\beta)\hat{\tilde{H}}_{HI}\hat{D}^{\dagger}(\beta)
\,.
\label{eq:h_hi}
\end{equation}
Likewise, we could keep the Hamiltonian fixed in its GCS form and instead transform the operators $\hat{O}$ and $\hat{H}_{ZE}$ as well as the initial state $\ket{\psi_{0}}$ from the QCS to the GCS. For technical reasons, however, we choose to transform the Hamiltonian, while keeping the operator and the initial state fixed for arbitrary $\beta$.

\section{Method}
\label{sec:method}

In order to numerically compute the time evolution in Eq. (\ref{eq:av_operator}), we need a basis to represent the state of our system and the operators acting on it. A natural choice for $N=1+K$ spins is given by the tensor product states of the electron spin and the nuclear spin eigenstates
\begin{equation}
\ket{n}
=
\ket{m^{n}_{S}}\otimes\bigotimes_{k=1}^{K}\ket{m^{n}_{k}}
=
\ket{\Downarrow\ua\da\da\ua\dots}
\,,
\label{eq:Hilbert_basis}
\end{equation}
where the electron spin is represented by $\ket{m^{n}_{S}}$, $m^{n}_{S}=\Downarrow,\Uparrow$ and the nuclear spin states by $\ket{m^{n}_{k}}$, $m^{n}_{k}=\da,\ua$. For convenience, we have ordered the nuclear spins $\ket{m^{n}_{S}m^{n}_{K}m^{n}_{K-1}\dots}$ according to the value of the envelope function at the corresponding site:
\begin{equation}
|\phi(r_{K})|^{2} \geq |\phi(r_{K-1})|^{2} \geq \dots
\end{equation}
Within this basis, an arbitrary state is given by a linear superposition of these $2^{N}$ states
\begin{equation}
\ket{\psi}
=
\sum\limits_{n=0}^{2^{N}-1} \alpha_{n} \ket{n}
\,,
\qquad
\sum\limits_{n=0}^{2^{N}-1} |\alpha_{n}|^{2}
=
1
\label{eq:state_superpos}
\end{equation}
with complex coefficients $\alpha_{n}$, while all operators are represented by $2^{N}\times2^{N}$ matrices. By diagonalizing the total Hamiltonian $\hat{M}\hat{H}\hat{M}^{\dagger}=\mathrm{diag}(\lambda_{0},\lambda_{1},\dots,\lambda_{2^{N}-1})$
we are able to re-express the time evolution operator
\begin{equation}
\hat{V}(t)
=
\hat{M}\hat{U}(t)\hat{M}^{\dagger}
=
\mathrm{diag}(\exp[-\I\hbar^{-1}\lambda_{0}\,t],\dots)
\,,
\label{eq:V_t}
\end{equation}
where $\hat{M}$ is an unitary operator formed by the eigenvectors of $\hat{H}$ and the $\lambda_{n}$ are the corresponding eigenvalues. Finally, we re-write Eq. (\ref{eq:av_operator}) and find
\begin{equation}
\av{O}(t)
=
\bra{\psi_{0}}\hat{M}^{\dagger}\;\;\hat{V}^{\dagger}(t)\;\;\hat{M}\,\hat{O}\,\hat{M}^{\dagger}\;\;\hat{V}(t)\;\;\hat{M}\ket{\psi_{0}}
\,,
\label{eq:av_operator_final}
\end{equation}
which we will evaluate for different parameter regimes in the following section. The numerical diagonalization is performed using the {\small\textrm{EIGEN}}\cite{Guennebaud2010} package for {\small\textrm{C++}}.

As one can notice from the explanations above, we deal with a quite big parameter space in which we can analyze the outcome of Eq. (\ref{eq:av_operator_final}). First, we control the shape of the dot by means of the envelope function $|\phi(r)|^{2}$, secondly the number of nuclear spins $K$ is variable and finally these spins can have different positions or configurations $C$ within the dot. All of these parameters change the AHI Hamiltonian in Eq. (\ref{eq:ahi}). Moreover, we will investigate different initial states $\ket{\psi_{0}}$ of the electron and the nuclear spins affecting Eq. (\ref{eq:av_operator_final}). Additionally, the eigenvector matrix $\hat{M}$, appearing in this equation is a function of the twisting angle $\beta$ between the GCS and the QCS. Note that the spectrum of eigenvalues $\lambda_{n}$ is unaffected by a change of $\beta$. Finally, we can also modify the absolute value of the external magnetic field, which we will parametrize by the resulting Zeeman energy of the electron $A_{ZE}$.

\begin{center}
  \begin{figure}
  \centering
  \includegraphics[width=0.45\textwidth,type=eps,ext=.eps,read=.eps]{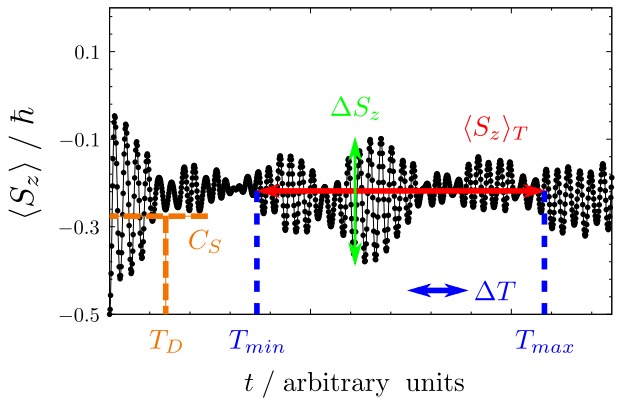}
  \caption{(Color online) Exemplary time evolution of the longitudinal electron spin component $\av{S_{z}}(t)$ as a function of time for out-of-plane orientation $\beta=0$, cf. Fig. \ref{fig:QD} (a). For a certain range of time $[T_{min},T_{max}]$ using a resolution $\Delta T$, we calculate the long-time average $\szlta$ and the standard deviation $\sigma_{S_{z}}$ and find the maximal deviation $\Delta S_{z}$ around this value. These quantities as well as the details of the oscillations including the beating structure depend on the choice of the parameters. The decoherence time $T_{D}$ is determined by a constant threshold $C_{S}$.}
  \label{fig:Sz_example}
  \end{figure}
\end{center}

\section{Results}
\label{sec:results}

In this section, we present our findings for the model system defined above. All calculations were carried out using an envelope function of the Gaussian type in Eq. (\ref{eq:envfunc}) with $R=7\;a_{NN}$ and a cut-off $C=10^{-6}$. This corresponds to a dot with diameter $D\approx7.2\,\mathrm{nm}$ containing $N_{sites}\approx10^3$ carbon atoms, such that $K=9$ atoms correspond to the natural abundance $n_{I}=0.01$ of $^{13}{\rm C}$. In order to investigate the impact of different initial states, we choose random complex (RC) initial states\cite{Schliemann2002,Schliemann2003}. These states were created by drawing complex coefficients $\alpha_{n}$ from $\mathrm{Re}[\alpha_{n}],\mathrm{Im}[\alpha_{n}]\in[-1,1]$ with equal probability and normalizing them according to Eq. (\ref{eq:state_superpos}). Moreover, we choose the electron spin always to point down resulting in initial states consisting of $\ket{\Downarrow\dots}$ states only, which means that $\alpha_{n}=0$ for $n\geq2^{K}$.

In order to determine qualitatively and quantitatively the impacts of the parameters, we investigate the time dependent expectation value $\av{S_{z}}(t)$ of the longitudinal electron spin component, which is calculated using Eq. (\ref{eq:av_operator_final}). A typical time evolution of $\av{S_{z}}(t)$ is plotted in Fig. \ref{fig:Sz_example}. Within the decoherence time $T_{D}$, the initial amplitude of the electron spin of $\hbar/2$ decays to its long-time average value, where still finite oscillations and beatings occur. This can be traced back to the finite size of the spin bath considered here.

Its long-time average value is calculated by
\begin{equation}
\szlta
=
\frac{1}{N_{T}}\sum\limits_{s=0}^{N_{T}}\av{S_{z}}(T_{min}+s\cdot\Delta T)
\,,
\label{eq:Sz_longtime}
\end{equation}
where we average over $N_{T}=(T_{max}-T_{min})/\Delta T$ time steps of width $\Delta T$. In order to investigate the oscillations of $\av{S_{z}}(t)$ quantitatively, we consider the standard deviation
\begin{equation}
\sigma_{S_{z}}
=
\sqrt{\frac{1}{N_{T}}\sum\limits_{s=0}^{N_{T}} \Big(\av{S_{z}}(T_{min}+s\cdot\Delta T) - \szlta\Big)^{2}}
\label{eq:Sz_stdv}
\end{equation}
as well as the sample range
\begin{equation}
\Delta S_{z}
=
\max_{t\in[T_{min},T_{max}]}[\av{S_{z}}(t)] - \min_{t\in[T_{min},T_{max}]}[\av{S_{z}}(t)]
\,.
\label{eq:Sz_delta}
\end{equation}
The latter quantity is a measure for the occurrence of oscillations with a big amplitude which originate from either recurrences of the signal, beatings or an entire lack of decoherence. While for beatings one expects rather small sample ranges $\Delta S_{z}<S_{z}(0)$, the former two cases should give values on the order of the initial amplitude, $\Delta S_{z}\sim O(S_{z}(0))$ in the out-of-plane case $\beta=0$ and  $\Delta S_{z}\sim 2 \cdot O(S_{z}(0))$ in the in-plane case $\beta=\pi/2$.

Besides these quantities characterizing the long time average of the electron spin, we are also interested in the amount of time it takes to decohere the system. In order to be independent from specific models of the decay, such as exponential or power-law decoherence, and to account for the characteristics of the numerics, we find this decoherence time $T_{D}$ by the first minimum exceeding a certain threshold $C_{S}$. For clarity, $C_{S}$ is also illustrated in Fig. \ref{fig:Sz_example}. This approach is similar to the one used in Ref. \onlinecite{Erbe2012} to find the decoherence times. Of course, the choice of this constant $C_{S}$ changes the value of $T_{D}$. However, its order of magnitude and its dependence on the different parameters is rather independent from a specific choice as long as $C_{S}$ is not too close to $\av{S_{z}}_{T}$, which we confirmed for different values of $C_{S}$.

In the following, we analyze both the decoherence time and the long-time average of the longitudinal electron spin component for different parameter sets. For each number $K$ of nuclear spins many initial states and configurations are created and labeled by numbers $0,1,2\dots$ for later comparison of the results. Note, that for different nuclear spin numbers $K$ these labels describe different initial states and configurations. Moreover, we concentrate on two orientations of the quantization axis, namely out-of-plane orientation for $\beta=0$ and in-plane orientation with $\beta=\frac{\pi}{2}$.

We investigated the effect of finite magnetic fields for exemplary initial states, configurations and $K=2,4,6$ nuclear spins, where we varied the resulting Zeeman constant from $A_{ZE}/A_{HI}\ll1$ to $A_{ZE}/A_{HI}\gg1$. For increasing $A_{ZE}$, we find a continuous crossover to a perfect alignment of the electron spin in the case of a very strong magnetic field. In the following, we put $A_{ZE}=0$ because we would like to better understand the low-magnetic field behavior of the spin dynamics in the presence of the AHI.

\subsection{Dependence of the long-time average on different initial states, configurations, and the number of nuclear spins}
\label{sec:sub:inst_config}

\begin{center}
  \begin{figure}
  \centering
  \includegraphics[width=0.45\textwidth,type=eps,ext=.eps,read=.eps]{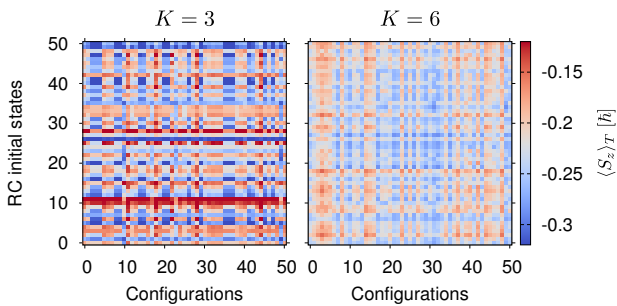}
  \caption{(Color online) Plot of the long-time average $\szlta$ for out-of-plane orientation $\beta=0$ and $K=3$ and $K=6$ nuclear spins without an external magnetic field. We considered $51$ different RC initial states and $51$ random configurations. For both numbers of nuclear spins, the electron spin looses roughly one half of its amplitude to $\szlta\approx-0.22\,\hbar$. The horizontal stripes for $K=3$ indicate the importance of the initial states for few nuclear spins, while configurations seem less relevant signaled by weaker vertical structures. This changes for $K=6$ nuclear spins, where the configurations dominate over initial states indicated by the vertical structures in the right plot.}
  \label{fig:Sz_mean__rd_050__c_050__A_+0.00e+00}
  \end{figure}
\end{center}

\begin{center}
  \begin{figure}
  \centering
  \includegraphics[width=0.45\textwidth,type=eps,ext=.eps,read=.eps]{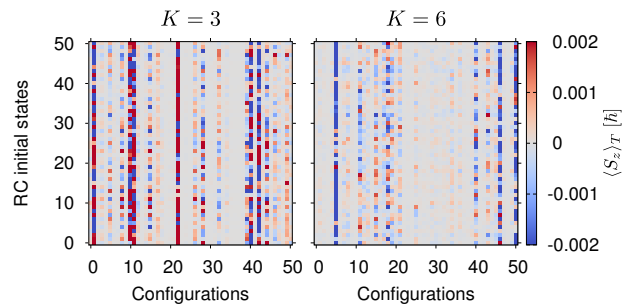}
  \caption{(Color online) Same plot as in Fig. \ref{fig:Sz_mean__rd_050__c_050__A_+0.00e+00} but for in-plane orientation $\beta=\pi/2$. For all parameters the long-time average of the longitudinal electron spin component sharply saturates at $\szlta\approx0$. In contrast to the out-of-plane case, the results seem independent of the initial state even in the few spin regime. For certain configurations, however, we find a non negligible dependence on the initial state.}
  \label{fig:Sz_mean__rd_050__c_050__A_+1.57e+00}
  \end{figure}
\end{center}

\begin{center}
  \begin{figure}
  \centering
  \includegraphics[width=0.45\textwidth,type=eps,ext=.eps,read=.eps]{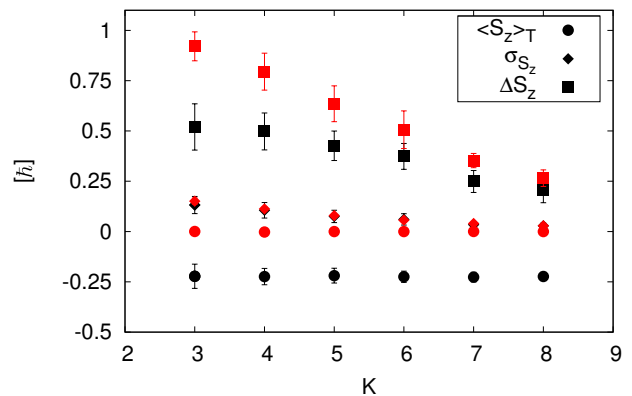}
  \caption{(Color online) Plot of the long-time average $\szlta$, the standard deviation $\sigma_{S_{z}}$ and, the sample range $\Delta S_{z}$ as a function of the number of nuclear spins $K$ for in-plane ($\beta=\pi/2$, red) and out-of-plane orientation ($\beta=0$, black). The values are obtained by averaging over $51$ RC initial states and $51$ different configurations, see Figs. \ref{fig:Sz_mean__rd_050__c_050__A_+0.00e+00} and \ref{fig:Sz_mean__rd_050__c_050__A_+1.57e+00}. Error bars are given by the standard deviation with respect to averaging over all $51\times51$ results. While the long-time average is almost constant, the averaged standard deviation $\sigma_{S_{z}}$ as well as the averaged sample range $\Delta S_{z}$ strongly decrease for larger $K$ indicating the reduction of fluctuations and of the occurrence of beating or recurrence events.}
  \label{fig:rd_C__C__rd_mean_sample_range}
  \end{figure}
\end{center}

First, we investigate the consequences of both different RC initial states and different configurations of the nuclei within the dot. We calculated $\szlta$, $\sigma_{S_{z}}$ and, $\Delta S_{z}$ for different parameter sets and found stable results for $T_{min}=0.5\times10^{9}\tau_{HI}$, $T_{max}=1.5\times10^{9}\tau_{HI}$, and $\Delta T=10^{4}\tau_{HI}$ with $\tau_{HI}=\hbar/A_{HI}\approx1\,\mathrm{ns}$. In Fig. \ref{fig:Sz_mean__rd_050__c_050__A_+0.00e+00}, we plot the long-time average $\szlta$ as a function of different RC states and configurations for $K=3$ and $K=6$, respectively, in out-of-plane orientation. The color map in Fig. \ref{fig:Sz_mean__rd_050__c_050__A_+1.57e+00} was created for the same parameters with in-plane orientation.

For a small number of nuclear spins $K=3$ and $\beta=0$, we observe strong fluctuations for both different $RC$ states and different configurations around an average value of $\szlta\approx-0.22\,\hbar$ as depicted in the color map of Fig. \ref{fig:Sz_mean__rd_050__c_050__A_+0.00e+00}. The horizontal stripes dominate over the vertical structures indicating, that the choice of the RC initial states has a greater influence on the results than the spatial configuration of the nuclear spins within the dot.

Moreover, we find large oscillations around this long-time average value for many configurations and initial states. This results in both sizable sample ranges $\Delta S_{z}$ and standard deviations $\sigma_{S_{z}}$. By averaging over all $51\times51$ results, we find $\av{\szlta}=(-0.22\pm0.06)\,\hbar$, $\av{\sigma_{S_{z}}}=(0.13\pm0.04)\,\hbar$, and $\av{\Delta S_{z}}=(0.52\pm0.11)\,\hbar$, which is also shown in Fig. \ref{fig:rd_C__C__rd_mean_sample_range}. The large average value of the sample range $\av{\Delta S_{z}}$ shows that for most cases analyzed, there was at least one big change in amplitude. However, no total spin flip to $+\hbar/2$ is achieved. The occurrence of sizable standard deviations indicates that there are on average many of these events. Thus in the few nuclear spin regime, coherent oscillations of the electron spin are the dominant dynamics, where recurrences of the initial value take place with a period $T_{P}=\hbar/\max_{i}(|\lambda_{i}|)\sim \hbar/(A_{HI}\cdot|\phi(r_{K})|^{2})\geq100\,\mathrm{ns}$.

If we consider a larger environment of nuclear spins as presented in Fig. \ref{fig:Sz_mean__rd_050__c_050__A_+0.00e+00} with $K=6$, the behavior of the long-time average changes. First of all, the result is much more uniform with respect to both the RC initial states and the configurations. In addition, the remaining differences in $\av{S_{z}}_{T}$ depend on the configurations rather than on the initial states, which is obvious from the vertical lines present in this color map. Averaging over all $51\times51$ results gives $\av{\szlta}=(-0.22\pm0.02)\,\hbar$, which is essentially the same as for $K=3$. However, the standard deviation $\av{\sigma_{S_{z}}}=(0.06\pm0.03)\,\hbar$ and the sample range $\Delta S_{z}=(0.37\pm0.06)\,\hbar$ clearly decrease. We confirmed this trend of decreasing fluctuations by repeating the above averaging procedure for other numbers of nuclear spins. These results are presented as a function of $K$ in Fig. \ref{fig:rd_C__C__rd_mean_sample_range}. While the long-time average value is constant, both the standard deviation and the sample range become smaller. Especially, the pronounced decay of the sample range clearly indicates that recurrences occur much less and, hence, that the corresponding recurrence times are increasing with more nuclear spins.

Thus, the major effect of an increased number of nuclear spins is to suppress the oscillations around the long-time average and changing the system from an initial state dominated regime to a regime where the configuration of the nuclear spins is important. This behavior can be understood by analyzing the impact of the nuclear spin number on the dimension of the Hilbert space and on the strength of the hyperfine interaction.

For a small number of nuclear spins, the dimension of the corresponding Hilbert space $D=2^{K+1}$ is small and, hence, we draw our RC initial states from a rather limited set, where individual single product states $\ket{n}$ lead to very different dynamics of the electron spin. Due to the combination of only $2^{K}$ states $\ket{n}$ to a RC initial state, it is not unlikely that one of these states dominates over the rest leading to rather diverse results.

By increasing K and, thus, the Hilbert space dimension this situation is changed. Since the individual state $\ket{\Downarrow\dots\ua\da}$ of nuclear spins at the border of the dot is almost irrelevant due to a small $|\phi(\vec{r}_{k})|^{2}$, groups of effectively equivalent states are superposed. Thus, a more effective averaging is achieved suppressing the dependence on a specific initial state. As a consequence, it is very unlikely for a single state to dominate over the rest.

The coupling strength of nuclei is the key in understanding the dependence of the results on the configuration. Its energy scale is given by the product $A_{HI}\,|\phi(r_{K})|^{2}$ of the hyperfine coupling constant and the maximal value of the envelope function at the sites of the nuclear spins. For a small number $K$, the probability to find two or more nuclear spins, which couple almost equally with the electron spin, is low due to the large gradient of the envelope function. Hence, effectively only one nuclear spin strongly interacts with the electron leading to simple oscillations. This behavior can be also easily derived by diagonalizing the resulting, effective $4\times4$ matrix of the AHI Hamiltonian given in Eq. (\ref{eq:ahi}). In doing so, one finds a discrete spectrum of frequencies given by the degenerate eigenvalues $\lbrace\lambda_{i}\rbrace=\lbrace-1/2,0,1/4,1/4\rbrace\cdot A_{HI}|\,\phi(r_{K})|^{2}$. This fact is responsible for the rather uniform dynamics with respect to different configurations in a small $K$ regime.

\begin{center}
  \begin{figure}
  \centering
  \includegraphics[width=0.45\textwidth,type=eps,ext=.eps,read=.eps]{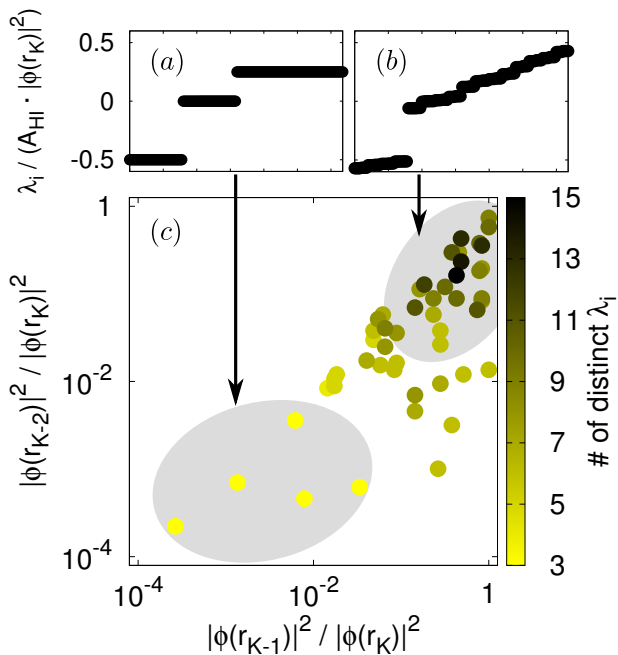}
  \caption{(Color online) $(a)$: Eigenvalues $\lambda_{i}$ of the AHI Hamiltonian for for $K=6$ nuclear spins and configuration $C=10$ in normalized units of $A_{HI}\,|\phi(r_{K})|^{2}$. $(b)$: Eigenvalues $\lambda_{i}$ for $K=6$ and $C=3$. $(c)$: Number of distinct eigenvalues $\lambda_{i}$ as a function of the relative probabilities $|\phi(r_{K-1})|^{2}/|\phi(r_{K})|^{2}$ and $|\phi(r_{K-2})|^{2}/|\phi(r_{K})|^{2}$ for $K=6$ nuclear spins. If both $|\phi(r_{K-1})|^{2}/\phi(r_{K})|^{2}\approx1$ and $|\phi(r_{K-2})|^{2}/\phi(r_{K})|^{2}\approx1$ at least three nuclear spins are strongly interacting with the electron spin causing a spectrum with many different eigenvalues as depicted in $(b)$. If only the most central nuclear spin couples strongly to the electron spin (lower left part), the spectrum is highly degenerate showing only three different eigenvalues as shown in $(a)$. The upper limit for the number of $15$ has no deeper meaning besides distinguishing both types of spectra.}
  \label{fig:N_007__zero_thd_+1.00e-04__jump_thd_+1.00e-01__phi1_phi2_ev_bin_num}
  \end{figure}
\end{center}

This situation can of course also occur for larger nuclear spin environments, as shown in Fig. \ref{fig:N_007__zero_thd_+1.00e-04__jump_thd_+1.00e-01__phi1_phi2_ev_bin_num} $(a)$ for $K=6$. It is, however, rather the exception from the more probable case of several nuclei coupling comparably to the electron, where a almost continuous spectrum is found as depicted in Fig. \ref{fig:N_007__zero_thd_+1.00e-04__jump_thd_+1.00e-01__phi1_phi2_ev_bin_num} $(b)$. If we characterize these spectra quantitatively by counting the number of distinct eigenvalues, i.e., eigenvalues which differ significantly, we can map the configuration of the nuclei to the spectra as depicted in Fig. \ref{fig:N_007__zero_thd_+1.00e-04__jump_thd_+1.00e-01__phi1_phi2_ev_bin_num} $(c)$.

For the in-plane case, our findings are quite different from the former ones. The electron spin saturates around $\szlta=0$ for both $K=3$ and $K=6$ as shown in Fig. \ref{fig:Sz_mean__rd_050__c_050__A_+1.57e+00}. Interestingly, we find already for $K=3$, that this average is reached very precisely with smaller fluctuations than in the out-of-plane case. This fact becomes also clear from averaging the longitudinal electron spin $\av{\av{S_{z}}_{T}}=(0.000\pm0.004)\,\hbar$ over all results. Moreover, the results are independent from the choice of the RC initial state. Some single configurations, however, give rise to deviations from this, where also a dependence on the initial state is restored. It seems, that this is the case, where several nuclear spins couple comparably to the electron spin explaining the sensitivity on initial states. The size of the fluctuations is on average given by $\av{\sigma_{S_{z}}}=(0.15\pm0.02)\,\hbar$. The mean value of the sample range of $\av{\Delta S_{z}}=(0.92\pm0.07)\,\hbar$ close to $1$ indicates, that in most cases, the electron spin is at least once almost completely flipped to $+\frac{1}{2}\,\hbar$ in contrast to the out-of-plane orientation. The $K=6$ study shows qualitatively the same result with $\av{\av{S_{z}}_{T}}=(0.000\pm0.001)\,\hbar$, where the fluctuations $\av{\sigma_{S_{z}}}=(0.057\pm0.004)\,\hbar$ are further suppressed. Moreover, the appearance of recurrences and total spin flips is also strongly decreased for $K=6$ as is clear from the sample range $\av{\Delta S_{z}}=(0.51\pm0.09)\,\hbar$. Analyzing this observable as a function of the number of nuclear spins, we observe again a prominent suppression of the fluctuations for growing $K$ as is apparent in Fig. \ref{fig:rd_C__C__rd_mean_sample_range}. 

In order to understand the differences between the in-plane and out-of-plane dynamics of the electron spin in more detail, an analytic analysis of the dynamics in the case of only one nuclear spin is very useful. Calculating the long-time average analytically for $K=1$ yields
\begin{align}
& \av{S_{z}}_{T}(\beta)
=
\lim_{\Delta T\rightarrow\infty}\frac{1}{2\Delta T}\int\limits_{T+\Delta T}^{T-\Delta T} \av{S_{z}}(t,\beta)
\nonumber\\
& =
-\frac{\hbar}{4}\cos(\beta)\Big[2\rho_{\da\da}\cos(\beta)+(\rho_{\ua\da}+\rho_{\da\ua})\sin(\beta)\Big]
\,,
\label{eq:szlta_analytically}
\end{align}
where the initial density matrix
%\
\begin{equation}
\rho_{0}
=\ket{\psi_{0}}\bra{\psi_{0}}
=
\left(
  \begin{array}{cccc}
  \rho_{\da\da}	& \rho_{\da\ua}	& 0	& 0	\\
  \rho_{\ua\da}	& \rho_{\ua\ua}	& 0	& 0	\\
  0		& 0		& 0	& 0	\\
  0		& 0		& 0	& 0	\\
  \end{array}
\right)
\label{eq:rho_0}
\end{equation}
is only non-zero for the electron spin pointing down as for the RC initial states. For more nuclear spins involved, the resulting equations become much more complicated. However, for the special case of only one strongly coupling nuclear spin, the structure of the AHI Hamiltonian remains the same and Eq. (\ref{eq:szlta_analytically}) still holds.

\begin{center}
  \begin{figure}
  \centering
  \includegraphics[width=0.45\textwidth,type=eps,ext=.eps,read=.eps]{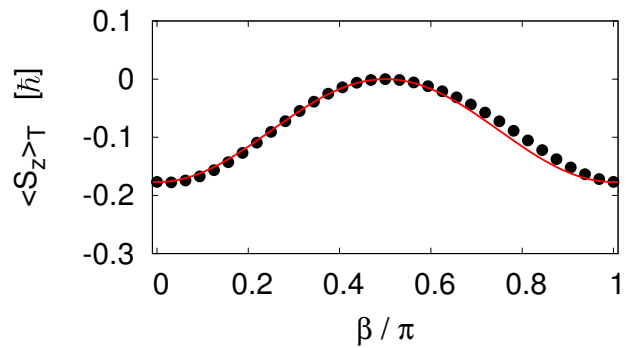}
  \caption{(Color online) Dependence of the long-time average $\av{S_{z}}_{T}(\beta)$ on the orientation of the quantization axis with respect to the graphene plane for $K=6$ nuclear spins. The example shown here was calculated for the configuration $C=3$, whose continuous spectrum is presented in Fig. \ref{fig:N_007__zero_thd_+1.00e-04__jump_thd_+1.00e-01__phi1_phi2_ev_bin_num} (b). The only parameter used to fit the numerical values to the analytic curve $\av{S_{z}}_{T}(\beta)=\av{S_{z}}_{T}(0)\cdot\cos^{2}(\beta)$ is the out-of-plane value $\av{S_{z}}_{T}(0)$.}
  \label{fig:beta_example}
  \end{figure}
\end{center}

We investigated the $\beta$ dependence of the long-time average numerically for some configurations and initial states and $K=2,4,6$ and $9$ nuclear spins, where we find good agreement of our results with $\av{S_{z}}_{T}(\beta)=\av{S_{z}}_{T}(0)\cos^{2}(\beta)$ with increasing $K$. Particularly, we observed this behavior also for configurations with several nuclear spins coupling almost equally to the electron spin. As an example, we plot in Fig. \ref{fig:beta_example} the $\beta$ dependence of $\av{S_{z}}_{T}$ for $K=6$ and configuration $C=3$. Its spectrum is shown in Fig. \ref{fig:N_007__zero_thd_+1.00e-04__jump_thd_+1.00e-01__phi1_phi2_ev_bin_num} (b). This fact is also supported by our results presented in Figs. \ref{fig:Sz_mean__rd_050__c_050__A_+0.00e+00} and \ref{fig:Sz_mean__rd_050__c_050__A_+1.57e+00}, where we find on average $\szlta\approx-0.22\,\hbar$ for $\beta=0$ and $\szlta\approx0$ for $\beta=\pi/2$. The deviation of $\av{S_{z}}_{T}(\beta=0)$ from $-\hbar/4$ originates from the finite time window $[T_{min},T_{max}]$ used in the numerical calculations, which misses recurrences of the full initial value of $\av{S_{z}}(t=0)=-\hbar/2$.

From this numerical findings and Eqs. (\ref{eq:szlta_analytically}) and (\ref{eq:rho_0}), we suppose that contributions from the off diagonal parts cancel each other almost completely and that the elements of the diagonal parts of the density matrix $\rho_{\da\da},\rho_{\ua\ua}$ have approximately equal weight of $1/2^{K}$, which seems reasonable for random complex initial states.

\subsection{Decoherence times}
\label{sec:sub:decoherence_times}

In this section, we want to investigate the decoherence times of the longitudinal electron spin $S_{z}$ for different initial states and different configurations. We chose the threshold to be always about $0.1\,\hbar$ below the obtained long-time average, which gives $C_{S}=-0.325\,\hbar$ for the out-of-plane case $\beta=0$ and $C_{S}=-0.1\,\hbar$ for the in-plane case $\beta=\pi/2$. Moreover, we used exactly the same initial states and configurations for all $K$ as for the calculation of the long-time average. The decoherence times were estimated for times up to $10^{7}\,\tau_{HI}\approx 10\,\mathrm{ms}$ with a time resolution $\Delta T = 10^{2}\,\tau_{HI}$, which yields at least $P=2\pi/(\Delta T \cdot \lambda_{max})\approx 20$ points per period of the highest absolute frequency $\max_{i}(|\lambda_{i}|)$. For $K=6$ we extended the investigated time regime to $10^{8}\,\tau_{HI}\approx 100\,\mathrm{ms}$ using the same time resolution $\Delta T$.

As it turns out, the decoherence times obtained by this method are rather independent from the initial states. Several factors are important for this fact. First of all, for larger numbers of nuclei of course the same arguments concerning the Hilbert space dimensions as for the long-time average hold. However, we also find for small $K$ only little dependence on the initial states. One reason for this is probably, that our method is robust against small changes of the longitudinal electron spin caused by different initial states, since we measure when the signal is above a certain threshold, but not how much. Finally, as we show below, the decoherence seems strongly related to the presence of many incommensurate frequencies. These frequencies are proportional to the eigenvalues of the hyperfine Hamiltonian and, hence, independent from the initial state.

Therefore, we focus in the following on the consequences of different configurations on the decoherence times for different numbers of nuclear spins. In principle, there are two relevant aspects concerning the positions of the nuclei, the absolute value of the envelope function $|\phi(\vec{r}_{K})|^{2}$ at the site of the strongest coupling nuclear spin and the relative position of the nuclei with respect to each other. The importance of the former is obvious, since the envelope function sets the maximal energy scale of the AHI in Eq. (\ref{eq:h_hi}) to $A_{HI}\cdot |\phi(r_{K})|^{2}$ and, consequently, rescales all times by a factor $|\phi(r_{K})|^{-2}$. Therefore, if we want to analyze the influence of the relative positions, we have normalize the decoherence times according to $T_{D}\rightarrow T_{D}\cdot|\phi(r_{K})|^{2}$.

We begin our discussion with investigating these normalized decoherence times for $K=6$ nuclei in more detail and then turn to absolute decoherence times as a function of $K$ afterwards.

\begin{center}
  \begin{figure}
  \centering
  \includegraphics[width=0.45\textwidth,type=eps,ext=.eps,read=.eps]{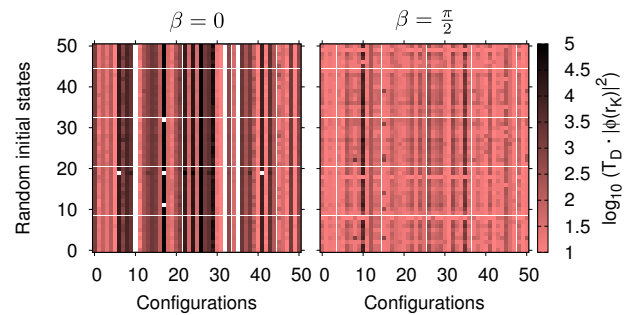}
  \caption{(Color online) Normalized decoherence time $T_{D}\cdot|\phi(r_{K})|^{2}$ as a function of $51$ RC initial states and $51$ random configurations for $K=6$ nuclear spins in in-plane and out-of-plane orientation. While the decoherence time is almost the same for different initial states, it strongly depends on the configurations showing deviations over several orders of magnitude. White spaces indicate the total lack of decoherence up to absolute times of $0.1\,\mathrm{s}$ given a threshold of $C_{S}=-0.325\,\hbar$. For special configurations, $C=10,32$ and $35$, and $\beta=0$ there is no decoherence at all, but coherent oscillations of the electron spin.}
  \label{fig:tr__N_007__A_+0.00e+00__THD_-3.25e-01__rd_map}
  \end{figure}
\end{center}

\begin{center}
  \begin{figure}
  \centering
  \includegraphics[width=0.45\textwidth,type=eps,ext=.eps,read=.eps]{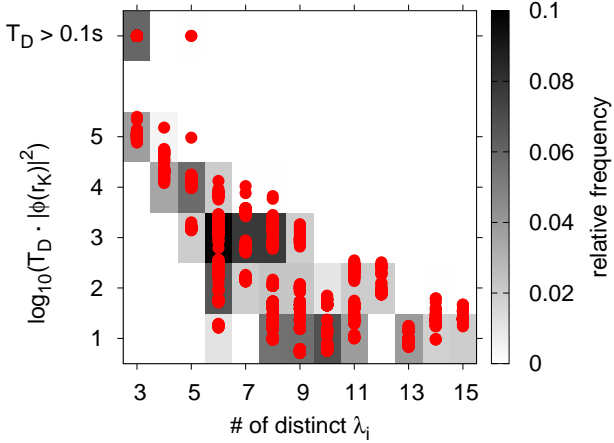}
  \caption{(Color online) Normalized decoherence time $T_{D}\cdot|\phi(r_{K})|^{2}$ as a function of the number of distinct eigenvalues of the AHI Hamiltonian for $K=6$ nuclear spins in out-of-plane orientation. For this plot all out-of-plane results presented in Fig. \ref{fig:tr__N_007__A_+0.00e+00__THD_-3.25e-01__rd_map} are considered. Obviously, long decoherence times occur only for a small number of distinct eigenvalues. Together with the results of Fig. \ref{fig:N_007__zero_thd_+1.00e-04__jump_thd_+1.00e-01__phi1_phi2_ev_bin_num} the importance of the relative coupling strengths $\propto|\phi(r_{K-1})|^{2}/|(\phi(r_{K}))|^{2}\;,\;\dots$ and, hence, of the relative position of different nuclei becomes evident.}
  \label{fig:N_007__zero_thd_+1.00e-04__jump_thd_+1.00e-01__A_+0.00e+00__THD_-3.25e-01_tr_bin_num_map}
  \end{figure}
\end{center}

A color map of the normalized decoherence times for $51$ initial states and $51$ configurations is shown in Fig. \ref{fig:tr__N_007__A_+0.00e+00__THD_-3.25e-01__rd_map}. For the out-of-plane case, we find that the decoherence times are almost independent of the initial state, but vary over several orders of magnitude for different configurations. If we plot the normalized times as a function of the number of distinct eigenvalues, cf. Fig. \ref{fig:N_007__zero_thd_+1.00e-04__jump_thd_+1.00e-01__phi1_phi2_ev_bin_num}, we find a direct connection between these times and the configuration of the nuclei in the dot. As is clear from Fig. \ref{fig:N_007__zero_thd_+1.00e-04__jump_thd_+1.00e-01__A_+0.00e+00__THD_-3.25e-01_tr_bin_num_map}, long decoherence times can be only found for the discrete spectra, which are realized if only one nuclear spin strongly interacts with the electron. The configurations without any decoherence, which are indicated by white spaces in Fig. \ref{fig:tr__N_007__A_+0.00e+00__THD_-3.25e-01__rd_map}, exhibit discrete spectra with the minimal number of distinct eigenvalues of $3$. An example of such a spectrum is shown in Fig. \ref{fig:N_007__zero_thd_+1.00e-04__jump_thd_+1.00e-01__phi1_phi2_ev_bin_num} (a). In these cases the dynamics of the longitudinal electron spin are coherent oscillations, where recurrences appear with a period of $T_{P}=|\phi(r_{K})|^{-2}\cdot\tau_{HI}\geq100\,\mathrm{ns}$. In contrast to this, short normalized decoherence times are a consequence of continuous spectra as presented in Fig. \ref{fig:N_007__zero_thd_+1.00e-04__jump_thd_+1.00e-01__phi1_phi2_ev_bin_num} (b). Thus, by the configurations studied, we can proof a direct relation between the relative positions of the nuclear spins and their relative coupling strengths, respectively, and the order of magnitude of the decoherence times.

For the in-plane case, the qualitative picture is similar, however, with shorter normalized decoherence times over all, such that we find decoherence within the investigated times for all configurations. In contrast to the out-of-plane case, also discrete spectra can show rather short decoherence times for specific configurations. Altogether, this demonstrates a much faster decoherence due to the broken symmetry in the in-plane orientation.

\begin{center}
  \begin{figure}
  \centering
  \includegraphics[width=0.45\textwidth,type=eps,ext=.eps,read=.eps]{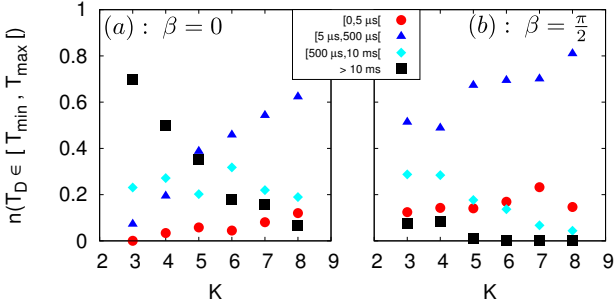}
  \caption{(Color online) Relative number of absolute decoherence times $T_{D}$ falling in a certain time interval $[T_{min},T_{max}[$ for different numbers $K$ of nuclei in in-plane and out-of-plane orientation. For each $K$ $51$ RC initial states and $51$ configurations were considered leading to $N_{calc}=2601$ calculations in total. $(a)$: In the out-of-plane case, long decoherence times are clearly dominating for few nuclear spins. Increasing $K$ leads to a quick decay of the decoherence times such that short times $T_{D}$ are common. Very short decoherence times start to become relevant for $K>6$. $(b)$: In the in-plane case, even for few nuclear spins short relaxation times are the rule. For increasing $K$, the percentage of short decoherence times is growing further.}
  \label{fig:gauss-----__tr_bins_oop}
  \end{figure}
\end{center}

Turning from normalized times to absolute decoherence times, the value of the envelope function $|\phi(r_{K})|^{2}$ at the site of the strongest coupling nuclear spins additionally becomes relevant, since it sets the order of magnitude of all times. Putting a larger and larger number of nuclear spins on a QD of constant area increases the average value of $|\phi(r_{K})|^{2}$, since it is more likely to find a spin very close to the center. Moreover, as we discussed above, an increased $K$ makes it much more probable to have several nuclear spins coupling almost equally to the electron spin. Altogether, this lets us expect a prominent decay of long decoherence times as a function of growing $K$, which is confirmed by Fig. \ref{fig:gauss-----__tr_bins_oop}. For $\beta=0$ and very few nuclear spins $K=3$, we find that the majority of decoherence times is longer than $10\,\mathrm{ms}$, whereas very short $T_{D}$ are almost completely irrelevant. For $K=8$ the percentages of short and long times are inverse. Now, only less than $7\%$ of the decoherence times are longer than $10\,\mathrm{ms}$, while most of the decay of the electron spins takes place within $500\,\mu\mathrm{s}$. However, for $K=6$, surprisingly, still about one-fifth of the cases shows ultra long decoherence times.

In the in-plane orientation, long decoherence times make up only a small fraction even for few nuclear spins. Short decoherence times in the range of $5\,\mu\mathrm{s}$ to $500\,\mu\mathrm{s}$ are significantly increasing for more spins. Notably, ultra short times below $5\,\mu\mathrm{s}$ do not become much more important.

In summary, typical decoherence times are on the order of $\mathrm{ms}$ under ideal conditions of small nuclear spin numbers and out-of-plane orientation. In the case of such long decoherence times, of course, other effects like spin orbit coupling could become relevant. In the presence of acoustic phonons and small external magnetic fields, this spin orbit coupling\cite{Struck2010,Hachiya2013} can lead to spin relaxation times of $T_{1}\sim1\,\mathrm{ms}$ below the decoherence times found here.

For larger numbers of nuclear spins and, generally, for in-plane orientation, decoherence times are smaller, but still above $5\,\mu\mathrm{s}$. Typical decoherence times of $GaAs$ QDs under spin echo\cite{Koppens2008} lie in the $T_{2,echo}\sim1\mu\mathrm{s}$ regime, whereas the current record of $T_{2,CPMG}\approx200\,\mu\mathrm{s}$ was measured using the Carr-Purcell-Meiboom-Gill (CPMG) pulse sequence\cite{Bluhm2010}. Pure dephasing times $T^{*}_{2}$ are below $50\,\mathrm{ns}$ for $GaAs$. Although all our estimates for the decoherence times are done for a model without any effort to improve the coherence of the electron spin like pulse sequences or strong magnetic fields, in almost all considered cases, we are above the $GaAs$ spin echo time $T_{2,echo}$. For smaller nuclear spin numbers, graphene even outperforms the CPMG time, which lets us expect very long decoherence times in graphene QDs when using pulse sequences.

\section{Summary and Conclusion}
\label{sec:conclusion}

Starting from a generic model of a graphene QD, we studied the dynamics of the electron spin caused by the hyperfine interaction with the nuclear spins present in the dot. The number of nuclei was varied from $K=2$ to $K=9$, where the upper limit corresponds to the natural abundance of spin carrying $^{13}{\rm C}$ for the dot size considered in this article. Besides the role of the number of nuclei, we also investigated the influence of the initial conditions as well as the impact of different configurations of the nuclei in the dot. Moreover, we explored the consequences of the orientation of the spin quantization axis with respect to the graphene plane. In order to characterize and quantify these effects, we analyzed both the long-time average $\av{S_{z}}_{T}$ of the longitudinal electron spin component and its decoherence time $T_{D}$.

Since nuclear spins are usually very hard to control in the envisioned experiments, we chose the initial states to be random complex (RC) superpositions of single product states. For this class of initial states, we found an appreciable effect on the long-time average only in the case of very few nuclear spins with $K<5$. Upon increasing the number of nuclear spins the effects of quantum parallelism and amplitude averaging\cite{Schliemann2002,Schliemann2003} reduce the differences between individual RC initial states more and more effectively. In this parameter regime, the results are dominated by the configuration of the nuclear spins within the dot, i.e., by their relative positions with respect to each other. For different configurations, the spectrum of eigenvalues of the hyperfine interaction varies from a highly discrete one with many degenerate eigenvalues to a continuous spectrum with many incommensurate frequencies.

For all $K$, a pronounced dependence of the long-time average on the orientation angle $\beta$ between the spin quantization axis and the normal vector of the graphene plane was found. It saturates at approximately one-half of its initial value of $\av{S_{z}}_{T}\approx-\hbar/4$ for $\beta=0$ and at $\av{S_{z}}_{T}\approx0$ in the in-plane case with $\beta=\pi/2$. While the long-time average of the electron spin is surprisingly almost constant with respect to $K$, we observed a strong reduction of fluctuations around it for larger nuclear spin baths.

In contrast to the long-time average, the decoherence times $T_{D}$ never showed a recognizable dependence on the initial states. Instead, the decoherence times depended decisively on the configuration of the nuclear spins in the dot. Long decoherence times were observed for only one nuclear spin strongly interacting with the electron spin, while several almost equally coupled nuclei lead to a very fast decoherence. Moreover, the decoherence times showed a strong dependence on the number of nuclear spins as well as on the orientation of the quantization axis. In the out-of-plane case, about $75\%$ of our results experienced decoherence times longer than $T_{max}=10\mathrm{ms}$ for $K=3$. For $K=8$, instead, less than $10\%$ showed no decoherence within this time frame while, in most cases, the electron spin decayed in less than $500\,\mu\mathrm{s}$. Considering the in-plane orientation, already for $K=3$ the majority of investigated initial state / configuration sets decohere within $500\,\mu\mathrm{s}$.

Although our results were obtained for a specific model of the graphene quantum dot using a Gaussian envelope function, they could be generalized quite naturally. In our model, the QD was comparably small with a sharp boundary. This choice resulted in a steep envelope function. Physically, this situation corresponds approximately to an etched QD. Thinking of larger QDs with smoother boundaries, we expect a flatter envelope function which gives rise to more nuclear spins interacting comparably with the electron spin. Consequently, it becomes more likely to end up with rather low fluctuations around the long-time average and to find quite short decoherence times. In contrast, the realization of even smaller dots\cite{Barreiro2012} with diameters of about $1\,\mathrm{nm}$ causes a very steep envelope function. This case should result in, at most, one nuclear spin interacting with the electron spin.

Both scenarios seem experimentally interesting in order to engineer QDs for different applications. A $^{13}{\rm C}$ enriched QD could potentially be used to prepare the electron spin very precisely in a certain superposition of spin up and down for subsequent experiments. A very small QD, in contrast, could serve as a storage for the electron spin where very long decoherence times are to be expected.

Besides these technical points, graphene QDs could also serve as a rich playground to test fundamental aspects of quantum mechanics and quantum information theory in an interesting system-bath setup. If we consider the hosted electron spin as the system, we are able to control both its spatial size and its state electrostatically. Moreover, the electron spin can be straightforwardly addressed via external magnetic fields or in an optical way. In contrast, a direct preparation of the state of the nuclear spin bath seems challenging. However, the size of the nuclear spin bath can be modified systematically by isotopic purification of either $^{12}{\rm C}$ or $^{13}{\rm C}$. Finally, as we argued above, the design of the QD enables the experimentalist to manipulate the strength as well as the nature of the system-bath interaction. Thus, these considerations render graphene QDs to be a flexible system-bath realization offering a controllable, fermionic bath of spin $1/2$ nuclei.

Given these opportunities, our setup not only seems very promising for studying a quantum to classical crossover as a function of the bath size in a fermionic environment, but also for more advanced concepts of quantum information theory such as quantum Darwinism\cite{Zurek2009}. With the notion of ``quantum Darwinism,'' W. H. Zurek summarizes his ideas of emergent classicality in a pure quantum universe, where the formation of classical and quantum system bath correlations, as well as the accompanied information exchange matter. Since measurements are most often indirect, relying on the environment as a mediator, it seems to be evident that the environment plays the crucial role in the creation of objective properties. However, as far as we know, this theory is up to now tested only in few experiments\cite{Burke2010,Cornelio2012} in a rather indirect way. In our opinion, taking advantage of the controllable spin bath in graphene QDs, this system offers a unique opportunity to reveal new insights in the role of the environment in the classical world we experience.

\section{Acknowledgements}
\label{sec:aknowledgement}

We would like to thank Manuel Schmidt for valuable discussions.
M.F. and B.T. acknowledge support from the Priority Program 1459 ``Graphene'' of the DFG and from the Euro-GRAPHENE Program of the ESF. The work of J.S. was supported by DFG via Collaborative Research Center 631. 

% \bibliographystyle{apsrev4-1}
% \bibliography{library.bib}

%Merlin.mbs v4.21 2009-07-09.
%

\end{document}